\documentclass[rmp,aps,preprintnofootinbib,endfloats]{revtex4}

\begin{document}

\def\ra{{\rightarrow}}
\def\a{{\alpha}}
\def\b{{\beta}}
\def\l{{\lambda}}
\def\eps{{\epsilon}}
\def\T{{\Theta}}
\def\t{{\theta}}
\def\co{{\cal O}}
\def\car{{\cal R}}
\def\caf{{\cal F}}
\def\cs{{\Theta_S}}
\def\pr{{\partial}}
\def\tri{{\triangle}}
\def\na{{\nabla }}
\def\S{{\Sigma}}
\def\s{{\sigma}}
\def\sp{\vspace{.15in}}
\def\hs{\hspace{.25in}}

\newcommand{\be}{\begin{equation}} \newcommand{\ee}{\end{equation}}
\newcommand{\bea}{\begin{eqnarray}}\newcommand{\eea}
{\end{eqnarray}}





\title{Tunneling between de Sitter and anti de Sitter black holes 
in a noncommutative ${\mathbf D_3}$-brane formalism}

\author{Supriya Kar}
\email{skar@ictp.trieste.it, skkar@physics.du.ac.in}
\affiliation{The Abdus Salam International Centre for Theoretical Physics,
Strada Costiera 11, Trieste, Italy}

\affiliation{ Department of Physics and Astrophysics,
University of Delhi, Delhi 110 007, India }

\begin{abstract}

We obtain dS and AdS generalized Reissner-Nordstrom like black hole geometries in a curved $D_3$-brane
frame-work, underlying a noncommutative gauge theory on the brane-world. The noncommutative scaling 
limit is explored to investigate a possible tunneling of an AdS vacuum in string theory to dS vacuum in 
its low energy gravity theory. The Hagedorn transition is invoked into its self-dual gauge theory to
decouple the gauge nonlinearity from the dS geometry, which in turn is shown to describe a pure dS vacuum.

\end{abstract}

\date{05 July 2006}
\maketitle





\section{Introduction}

In the recent years, considerable amount of interest has been devoted to explore the possibility of
de Sitter (dS) vacua in quantum gravity \cite{witten1}-\cite{strominger2}. Contrary to the well understood
anti de Sitter (AdS) spaces \cite{maldacena}, the dS geometries are usually hard to perceive in a quantum theory.
The primary reason lies in the quantum tunneling of dS to AdS, which assures 
metastable dS vacua. The fact that the complete event horizon in an hyperbolic geometry is not accessible 
to an observer make dS in a different footing than AdS and Minkowski vacua. Interestingly, the construction 
of dS vacua has been achieved by taking into account a small number of $D_3$-branes along with the AdS vacua 
in a type IIB string theory \cite{kklt}.

\par
\sp
Among the recent developments, the nonlinear electromagnetic (EM-) field on a $D_3$-brane turns out
to be a potential candidate to address some of the quantum aspects of gravity 
\cite{gibbons-2}. In fact, consistent noncommutative deformations of Einstein gravity has been the subject of interest in the recent literature \cite{vassilevich}. In the context, a very recent reveiew may be found in
ref.\cite{sazbo}.

\par
\sp
In this paper, we obtain generalized $dS_4$ and $AdS_4$ Reissner-Nordstrom (RN-) like black hole geometries 
in a curved $D_3$-brane frame-work \cite{km-2}, underlying the noncommutative gauge theory on the brane-world
\cite{seiberg-witten}. We investigate the gravity decoupling regime initiated by the Hawking radiation phenomenon
from the black holes. A noncommutative scaling \cite{verlinde2} limit generated in the frame-work,
is explored to obtain the low energy gravity regime.
A priori, the theory may be seen to describe $2D$ extremal dS black hole geometry, which may 
alternately be viewed as a combination of AdS and dS geometries. 
However, the presence of three extra large dimensions in the regime is argued to elevate a near
horizon $dS_2$ geometry to an appropriate $dS_5$. The Hagedorn phase in the self-dual gauge-string theory 
is exploited to show that the extremal black hole Hawking radiates the EM-nonlinearity
\cite{kar-panda} and is described by a pure dS space. The analysis incorporates
a series of tunnelings among AdS and dS vacua and may provide a clue to our present day
metastable brane-world.

\section{CURVED ${\mathbf D_3}$-BRANE AND SMALL ${\mathbf\Lambda}$}

A $D$-brane governs the boundary $\pr{\cal M}$ dynamics of an open string.  The induced fields on the brane ($g_{\mu\nu}$ and antisymmetric $b_{\mu\nu}$) are the pull-back of the respective dynamical background fields
in the string bulk. In principle, the gravity dynamics can be incorporated into a curved brane frame-work along with the gauge dynamics of a $D_3$-brane. The formulation inspires one to seek for a fundamental theory \cite{vafa}
in presence of a three brane, such as D=12 constructions \cite{tseytlin}.
However our starting point, in this paper, lies in the bosonic sector of  $D=10$ type IIB string theory on $K_3\times T^2$. Ignoring the Chern-Simons terms, the relevant $4D$ effective string dynamics 
in Einstein frame may be given by
\be
S_{\rm string}= - \int d^4x \ {\sqrt{{\tilde G}^E}} \ \bigg ( {1\over{16\pi G_N}} R- 2(\pr\phi)^2
-{1\over{2}} F_1^{(k)} C_{kl} F_1^{(l)} - {1\over{2\cdot 2!}} F_2^{(i)}D_{ij}F_2^{(j)}\
-\ {{Z}\over{2\cdot 4!}} F_4^{(m)}L_{mn}F_4^{(n)} \ \bigg )\ ,\label{db1}
\ee
where ($C_{kl}$, $D_{ij}$,$L_{mn}$) govern the appropriate moduli coupling to the gauge field of various ranks
and $Z$ is a normalization constant. Then, the four form energy density becomes nontrivial and can be given by
a potential in the moduli space
\be
V_4(\phi) ={{Z}\over{48}}F_4^{(m)}L_{mn}F_4^{(n)}\ .\label{db2}
\ee
On the other hand, the $D_3$-brane dynamics has been worked out, explicitly, for constant induced fields only.
The Minkowski inequality in the theory enforces a self-duality of the EM-fields, in the $D_3$-brane dynamics. 
Then, the noncommutative gauge theory on the $D_3$-brane can be approximated by the Dirac-Born-Infeld dynamics.
It is given by
\be
S_{D_3}= -\int_{\pr{\cal M}} d^4x\ \sqrt{G} \left (\ \lambda_b  - {1\over4} G^{\mu\lambda}
G^{\nu\rho}\ {\hat F}_{\mu\nu}\star {\hat F}_{\lambda\rho} \ \right )\ ,\label{db3}
\ee
where $\lambda_b$ is the brane tension and $G \equiv \det G_{\mu\nu}$. The Moyal $\star$-product accounts 
for the nonlocality arise due to the infinite number of derivatives there. Importantly, the gravitational 
back reaction has been incorporated into the effective theory, which is apparent from the definition of 
the modified metric $G_{\mu\nu} = (g_{\mu\nu} - [ b g^{-1} b ]_{\mu\nu} + [ b g^{-1}b\ bg^{-1} b]_{\mu\nu} 
+ \dots )$. Now, the curved $D_3$-brane dynamics is obtained by coupling the noncommutative $D_3$-brane 
(\ref{db3}) to an effective string theory (\ref{db1}).
In a static gauge, the complete dynamics of a curved $D_3$-brane can be given by
\be
S= - \int d^4x\ {\sqrt{G}}\ \left (\ {1\over{16\pi G_N}}(R - 2\Lambda ) -2(\pr\phi)^2 -
{1\over2} F_1^{(k)}C_{kl}F_1^{(l)}
-{1\over4} F_2^{(p)} D_{pq}F_2^{(q)} \right )\ ,\label{db6}
\ee
\be
{\rm where}\qquad\quad \Lambda (\phi) = 8\pi G_N \left (V_4(\phi) - \lambda_b \right )\ .
\qquad\qquad\qquad\qquad\qquad\qquad\qquad {}\label{db7}
\ee
$\lambda_b$ can take a large constant value as it can be seen to be controlled by an $U(1)$ gauge 
non-linearity in the theory.
The multiple four forms in the theory together with the brane tension, 
redefine the vaccum energy (\ref{db7}). Since an explicit membrane dynamics is absent in the frame-work
(\ref{db6}), the (multiple) four form equations of motion are worked out to yield
$\pr^{\mu}\left ( \sqrt{G}\ L_{mn} F^{(n)}_{\mu\nu\lambda\rho}\right ) = 0$. 
For a stable minima in $V_4(\phi)$, the $L_{mn}$ takes a constant value. 
Then, the solution(s) to the equation(s) of motion are
given by $F^{(n)}_{\mu\nu\lambda\rho} = \lambda^{(n)}\eps_{\mu\nu\lambda\rho}$, where $\lambda^{(n)}$ are constants and $\eps_{\mu\nu\lambda\rho}$ is a totally antisymmetric tensor. Thus at a local minima,
the $\Lambda(\phi)$ takes a constant value
\be
\Lambda(\phi)\rightarrow 8\pi G_N 
\left ( {{Z}\over2} \sum_{n'=1}^{n} \left [\lambda^{(n')}\right ]^2\ -\ \lambda_b\right ) .\label{db8}
\ee
It implies that the multiple four-forms along with the
gauge non-linearity could possibly reduce the effective cosmological constant (\ref{db8}) 
to a small value in $4D$, which lies along the idea of dynamical neutralization \cite{brown-teitelboim}.
The potential, between moduli and second rank gauge fields in (\ref{db6}), becomes
$V_2(\phi)=\ - [{\hat Q}_{\rm eff}^2 + Q^{(i)}D_{ij}Q^{(j)}]$, where ${\hat Q}_{\rm eff}$ and $Q^{(i)}$
denote the electric (or magnetic) charges, respectively, on the brane and in the effective string theory.

\par
\sp
With a gauge choice $G_{i\a}=0$, for $(\a,\b)\equiv (x^4,x^1)$ and $(i,j)\equiv(x^2,x^3)$, the action (\ref{db6}) is
simplified using a noncommutative scaling \cite{verlinde2}. The scaling incorporates vacuum field
configurations for some of the field components:
$\partial_{\alpha}h_{ij}= 0$, $R_{\bar h} = 0$, $\pr_{\a}\varphi^{(m)}=0$ and $F^{(p)}_{\a\b}=0$.
Then, the relevant curved brane dynamics can be governed by its on-shell action. It is given by
\bea
&&S = -\int d^2x^{(\a)} d^2x^{(i)}
{\sqrt{\bar h}} {\sqrt{h}}\bigg [ {1\over{16\pi}}\left ( R_h - 2\Lambda\right )+
{1\over{64\pi}}h^{ij}\partial_i {\bar h}_{\alpha\beta}\partial_j{\bar h}_{\gamma\delta}
\epsilon^{\alpha\gamma}\epsilon^{\beta\delta}
\qquad\qquad\qquad\qquad {} \nonumber\\
&&\qquad\qquad\qquad\qquad\qquad\qquad\qquad\qquad\qquad\quad\;\
-2 h^{ij}C_{mn}\pr_i\varphi^{(m)} \pr_j\varphi^{(n)}
- {1\over2} {\bar h}^{\alpha\beta} h^{ij} D_{pq}{\hat F}^{(p)}_{\alpha i}\star{\hat F}^{(q)}_{\beta j}
\bigg ]\ ,\label{db142}
\eea
where $C_{mn}$ and $D_{pq}$ for $p,q=(1,2,\ \dots i, i+1)$ are the appropriate moduli couplings and $\varphi^{(m)}$ take into account the
dilaton and axions in the theory.
A most general static, spherically symmetric ansatz, for the metric in the frame-work is given by
\be
ds^2 =\ f \ dt_E^2\ +\ f^{-1} \ dr^2\ + \ h^2\ d\Omega^2 \ ,\label{db11}
\ee
where $f$ and $h^2$ are arbitrary functions of $r$.

\section{ ${\mathbf{dS}}$ and  ${\mathbf{AdS}}$  black holes}

\subsection{Constant moduli}

We consider constant moduli in the theory (\ref{db142}) and restrict the EM-field on the brane only, 
$i.e.\ {\hat Q}\neq 0$ and $Q^{(i)}=0$. The anstaz for ${\hat A}_{\mu}\equiv ({\hat A}_t, {\hat A}_r, 0, 0)$ 
becomes ${\hat A}_t = - {\hat Q}_{\rm eff} \sin\theta \cos\phi$ and ${\hat A}_r= {\hat Q}_{\rm eff} \sin\theta\sin\phi$. The non-vanishing components of the self-dual EM-field are
$E_{\theta}= B_{\theta} = ({\hat Q}_{\rm eff}/r) \cos \phi$  and $E_{\phi}= B_{\phi}= - ({\hat Q}_{\rm eff}/r)
\sin \phi$. Then, the independent components of Ricci tensor in the theory can be expressed in terms of $f(r)$, $h(r)$ and $V_2(\phi)$. The metric components are worked out to yield
\be
f_{\pm}=\ \left (1 \mp {{r^2}\over{b^2}} - {{2M_{\rm eff}}\over{r}}\right ) \left ( 1 \pm E^2 \right)
=\left ( 1 -{{2 M_{\Theta}}\over{r}} - {{\Lambda}\over3}r^2 \right)
\;\ {\rm and}\;\ h(r)= \left ( r^2 -{{2M_{\rm eff}{\hat Q}_{\rm eff}^2}\over{r}}\right )^{1/2},\label{db13}
\ee
where $f_+$ and $f_-$ signify the appropriate geometries $f(r)$, respectively, for the $dS$ and $AdS$ spaces.
The effective mass parameter $M_{\Theta}$ takes into account the noncommutative $\Theta$-corrections, from the
boundary string dynamics \cite{km-2}, to the ADM mass and charge of a black hole. Explicitly, it can be expressed as
\be
M_{\Theta}=M_0 \left [ 1 - {{\Theta}\over{2r^2}} + {\cal O}(\Theta^2) + \dots \right ]\ = \left ( 
M_{\rm eff} \pm  {{{\hat Q}_{\rm eff}^2}\over{2r}}\right )\ ,\;\;
{\rm where}\quad M_0=G_N\left ( M + {{{\hat Q}^2}\over{2r}}\right )
\ .\label{db131}
\ee
In the case, $M_{\rm eff}$ and ${\hat Q}_{\rm eff}$, respectively, denote the ADM mass and the
charge of a generalized black hole. To order ${\cal O}(G_N)$, the explicit geometry 
corresponding to dS and AdS RN-like black holes are given by
\be
ds^2 =\ -\left (1 \mp {{r^2}\over{b^2}} - {{2M_{\rm eff}}\over{r}} \pm {{{\hat Q}_{\rm eff}^2}\over{r^2}} \right) dt^2 + \left ( 1 \mp {{r^2}\over{b^2}} -{{2M_{\rm eff}}\over{r}} \pm {{{\hat Q}_{\rm eff}^2}\over{r^2}} \right)^{-1} dr^2 + r^2\ d\Omega^2\ ,\label{db14}
\ee
where $b$ is the dS (or AdS) radius as appropriate to a geometry. 
The generalized black hole geometries are characterized by three parameters ($\Lambda$, $M_{\rm eff}$ and ${\hat Q}_{\rm eff}$). The horizon equation $f(r)=0$ can be solved 
to obtain three physical horizons in $dS_2\times S^2$. In the
decreasing order of their radius, they are characterized by a cosmological horizon $r_c$, an
event horizon $r_+$ and an inner horizon $r_-$. Interestingly, for ($M=0={\hat Q}$) and 
$\Lambda\neq 0$, the black hole geometry reduces to a pure dS with a horizon at $r_c=b$.
Similarly, for ($M\neq 0, {\hat Q}\neq 0$) and $\Lambda=0$, the geometry corresponds to a 
generalized dS RN-like black hole with horizons at
$r^{\rm dS}_{\pm} = (M_{\rm eff} \pm [{M^2_{\rm eff} -{\hat Q}^2_{\rm eff}}]^{1/2})$.
On the other hand, the AdS radius incorporates a periodicity in time coordinate $t_E\rightarrow t_E+2\pi b$. 
For $\Lambda=0$, there is only one event horizon which is unlike to that of dS black hole. 
The radius of the event horizon though resembles to that of a typical Schwarzschild black hole $r_h^{\rm AdS} \simeq 2M_{\rm eff}$, it governs a regular geometry there.

\subsection{Arbitrary moduli}

A nonconstant moduli in the theory retains nontriviality in the effective potential.
In addition to the nonlinear $U(1)$ gauge potential on the brane, there are non-vanishing 
multiple $U(1)$ potentials in the case. We consider an appropriate anstaz for the multiple gauge 
fields $A^{(i)}$. The non-vanishing components of $A^{(i)}$ are the $\perp$-components and we consider them as
$A^{(i)}_t= Q^{(i)}/r$ and $A^{(i)}_{\phi}= Q^{(i)} \cos\theta$.
The corresponding EM-field(s) are given by
$F^{(i)} = Q^{(i)}\left [({1/{r^2}})\ dt\wedge dr + \sin\theta \ d\theta\wedge d\phi\right ]$.
The non-vanishing electric or magnetic field components are given by $E_r^{(i)}=B_r^{(i)} 
= Q^{(i)}/r^2$. Interestingly, with an orthogonal rotation, the arbitrary function $f(r)$ can be represented by
eq.(\ref{db13}). However the $E^2$ there, receives correction due to the multiple gauge fields at 
${\cal O}(G^2_N)$. It becomes
\be
E^2=\ {1\over{r^2}}\left ( {\hat Q}^2_{\rm eff}\ +\ {{G^2_N}\over{r^2}} Q^{(i)}D_{ij}Q^{(j)}\right )\ .
\label{db16}
\ee
On the other hand, the moduli significantly modifies the radius of $S^2$. It is computed to yield
\be
h =\ r \ e^{\phi(r)}\ ,\quad {\rm where}\quad e^{\phi(r)}= \left ( e^{2\phi_h} -
{{G_N}\over{r r_h}} Q^{(i)}D_{ij}Q^{(j)}\right )^{1/2}\ ,\label{db18}
\ee
where the constant $\phi_h$ is the value of $\phi$ at the event horizon $r_h$.
The effective mass $M_{\Theta}$, in the case, can be seen to accommodate higher order terms ${\cal O}(G_N^2)$.
Explicitly, the $G^{tt}$ component is given by
\be
f_{\pm} =\ \left ( 1 \mp {{r^2}\over{b^2}} - {{2M_{\rm eff}}\over{r}} 
\pm {{{\hat Q}_{\rm eff}^2}\over{r^2}}\right )
\ \pm\ {{G^2_N}\over{r^4}} Q^{(i)}D_{ij}Q^{(j)} \left (1-{{r^2}\over{b^2}}\right )
+ \ {\cal O}(G^3_N)\ .\label{db181}
\ee
Then, the generic dS and AdS RN-like black hole geometries, to ${\cal O}(G_N)$ in the theory, are given by
\bea
&&ds^2 = - \left ( 1 \mp {{r^2}\over{b^2}}- {{2M_{\rm eff}}\over{r}} \pm {{G_N{\hat Q}^2}\over{r^2}} \mp
{{\Theta {\hat Q}_{\rm eff}^2}\over{r^4}}\right ) dt^2\qquad\qquad\qquad {}\nonumber\\
&&\qquad\qquad\qquad\qquad
+ \left ( 1\mp {{r^2}\over{b^2}} - {{2M_{\rm eff}}\over{r}} \pm {{{G_N}{\hat Q}^2}\over{r^2}} \mp
{{\Theta {\hat Q}_{\rm eff}^2}\over{r^4}}\right )^{-1} dr^2\ +\  e^{2\phi(r)} \ r^2 d\Omega^2\ .\label{db19}
\eea
It implies that the area of the event horizon tend to shrink in presence of moduli in the theory.
Unlike to the $dS$ black hole, the $AdS$ geometry possesses only one horizon.
Our analysis suggests that the moduli corrections along ($t,r$)-space to 
$dS$ and $AdS$ black hole geometries begin at ${\cal O}(G^2_N)$. However, the shrinking radius of the
event horizon is reconfirmed even at ${\cal O}(G_N)$. In absence of $D_3$-brane and with $\Lambda=0$,
the $dS$ geometry reduces to the one obtained in an effective string theory \cite{garfinkle}.

\section{TUNNELING: ${\mathbf{AdS\leftrightarrow dS}}$}

\subsection{Gravity decoupling limit}

In presence of nonlinear EM-charges, $i.e.\ \Theta$ corrections, the interesting feature of the
gravity decoupling limit $g\rightarrow 0$ can be exhibited in the formalism. The limit describes the low energy
aspects of an effective curved brane and essentially governs a semi-classical regime. It 
can be checked that the usual extremal limit $M\rightarrow {\hat Q}$ can be reached by taking 
$M\rightarrow 0$ instead. In the limit, the generic $dS_4$ black hole (\ref{db19}) is governed by $dS_2\times S^2$.
The extremal dS and AdS black hole geometries are given by
\be
ds^2= - \left (1 \mp {{r^2}\over{b^2}}  \pm {{{\hat Q}_{\rm eff}^2}\over{r^2}} \right) dt^2 +
\left ( 1 \mp {{r^2}\over{b^2}} \pm {{{\hat Q}_{\rm eff}^2}\over{r^2}} \right)^{-1} dr^2 +
\left [r^2_h\ e^{2\phi_h} - {G_N} Q^{(i)}D_{ij}Q^{(j)}\right ] d\Omega^2\ ,\label{db21}
\ee
where $r_h$ is the radius of $S^2$ at the event horizon in absence of moduli.
It implies that the moduli at the event horizon can be expressed in terms of the $U(1)$ gauge charges, which in turn
can shrink the effective event horizon radius to a small value in the regime. On the other hand $f(r)=0$ relates
the effective event horizon radius to the nonlinear $U(1)$ charges. The radius of $S^2$ can be checked to
satisfy
\be
r_h^2 e^{2\phi_h} =\  {{b^2}\over2} \left [ \pm 1 + \left ( 1+ {{4{\hat Q}_{\rm eff}^2}\over{b^2}}\right )^{1/2}\right ] +\ {G_N} Q^{(i)}D_{ij}Q^{(j)} \ .\label{db22}
\ee
For a fixed ${\hat Q}_{\rm eff}$, the radius square of $S^2$ becomes $[b^2 -V_2(\phi_h)]$, for the 
dS geometry and $[-V_2(\phi_h)]$, for the AdS there. It implies that the effective radius of the event horizon is
not fixed in general, rather it is governed by $V_2(\phi)$ in the moduli space. In the decoupling regime, 
the moduli moves to its local minima along the potential, $i.e.\ V_2(\phi)\rightarrow V_2(\phi_h)$,
and hence decouples the $S^2$ from the effective $4D$ geometry.
The emerging $2D$ large black hole geometry is given by
\be
ds^2 = - r^2 dt^2 +\ {{dr^2}\over{r^2}}\ .\label{db23}
\ee
Naively, $dS_4$ and $AdS_4$ black holes appear to govern different geometries. However, in the
gravity decoupling limit they can be argued to overlap and describe a new $dS$ black hole in two dimensions. 
In particular, the event horizon $r_{\tilde h}$ of the $AdS$ describes the curvature singularity of 
the new dS black hole.

\subsection{Extra dimensions and $\Theta$-decoupling}

In the extremal limit, the generic black hole solutions (\ref{db19}) can be seen to
provide hint for the existence of three large extra dimensions.
In fact, the nonlinear EM-charges remain in the gravity decoupled regime govern an 
effective theory of gravity there. Naively, ${\hat Q}^2_{\rm eff}$ can 
be seen to correspond to a light mass for the
black hole in the effective theory. The fact that $1/r^2$ term is associated with a mass term
in the extremal black hole geometry (\ref{db19}))
favors the assertion of three extra large dimensions in addition to the 
$2D$ black hole geometry within the curved $D_3$-brane frame-work. In addition, the assertion can
further be re-confirmed in presence of $\Theta$-terms there in $dS$ and $AdS$ solutions. 
In particular, its association with the
$1/r^4$ term, a priori, predicts some appropriate seven dimensional theory. However, two of
the noncommutative constraints from the boundary theory, make the effective space-time dimension to five.
In other words the formulation urges for an underlying $5D$ effective theory of gravity instead of that in $4D$.
Very recently in a collaboration \cite{km-2}, the presence of a fifth dimension in the extremal limit has been exploited by using a noncommutative scaling  on the brane. The orthogonality in
coordinates make the fifth dimension $\perp$- to a generic $D_3$-brane world-volume. In otherwords, the
curvature in the curved brane theory becomes appreciable along the extra dimensions. Incorporating the required
large extra dimensions into the curved $D_3$-brane formalism, one may alternately view the 
$2D$ extremal black hole (\ref{db23}), either as an $dS_2\times S^3$ or as an $AdS_2\times S^3$ geometry. 
The result is in agreement with the $dS_5$/CFT \cite{strominger2} and precisely
with the $AdS_5$/CFT correspondence in string theory \cite{maldacena}.

\par
\sp
Now, the gauge theory perspective of extremal dS black hole (\ref{db19})
is investigated for its near horizon geometry. The Hawking radiation leading to
an extremal geometry essentially deescribes a typical $D_3$-brane, which corresponds to the event 
horizon of an $dS_5$ black hole. It can be given by
\be
ds^2= - \left (1- {{r^2}\over{b^2}}  + {{G_N{\hat Q}^2}\over{r^2}} - 
{{\Theta {\hat Q}^2_{\rm eff}}\over{r^4}} \right) dt^2 +
\left ( 1- {{r^2}\over{b^2}} + {{{\hat Q}_{\rm eff}^2}\over{r^2}} -
{{\Theta {\hat Q}^2_{\rm eff}}\over{r^4}}\right)^{-1} dr^2 + r^2\ d\Omega_3^2\ .\label{db26}
\ee
Here $r$ is the radius of $S^3$. The event horizon of $dS_5$ RN-black hole is at
$r_h= (b^2 + {\hat Q}^2_{\rm eff})^{1/2}$ and the curvature singularity is at 
$r= {{\hat Q}_{\rm eff}}$. Due to the non-trivial $U(1)$ charges
in the solutions, they are often referred as monopole black hole in literature \cite{gibbons-townsend}.
However, the no hair conjecture \cite{bizon} for the nonlinear charged black hole, make it unstable.
In other words, the extremal black hole undergoes Hawking radiation to decouple the $\Theta$-terms there. 
The nonlinear gauge decoupling can be seen to describe a second order phase transitions and
leads to the Hagedron phase in the frame-work \cite{kar-panda}. 
In the regime, the critical phase can be described by $(E^+_c - E^-_c)$, where the nonzero string
modes in the regime undergo an exchange between its real and imaginary components. As pointed out, it 
results in the decoupling of nonlinear gauge charges and lead to a RN-black hole solution in $5D$. Finally, 
the Hawking radiation ceases with a stable remnant of $U(1)$ gauge charge and the geometry may be described by
\be
dS^2 = - \left ( 1 - {{r^2}\over{b^2}} + {{G_N{\hat Q}^2}\over{r^2}} \right ) dt^2 +
\left ( 1 - {{r^2}\over{b^2}} + {{G_N{\hat Q}^2}\over{r^2}} \right )^{-1} dr^2 +\  r^2\ d\Omega_3^2
\ ,\label{db27}
\ee
where the radius of the event horizon is $r_h= b$ and  the curvature singularity
can be seen to be at $r=0$. The monopole black hole solution seems to possess two interesting features. 
For large $r$, it reduces to a pure $dS_5$ geometry, possibly describing our $4D$ brane-world on its boundary
\cite{strominger2}. On the other hand, at Planck scale the dS black hole reduces to a precise monopole solution with
asymptotically flat geometry, $i.e. \Lambda=0$. It possiblly re-assures our earlier assertion that a nonlinear EM-field gives rise to a cosmological constant in a curved brane theory.

\section{CONCLUDING REMARKS}

To conclude, dS and AdS (generalized) RN-like black hole geometries were obtained in a curved $D_3$ frame-work
underlying a noncommutative $U(1)$ gauge theory on its brane-world. The small value of cosmological constant
was argued in the frame-work following a dynamical neutralization technique. The frame-work was shown to
accommodate multiple $U(1)$ gauge fields coupled to moduli, which are in addition to the noncommutative gauge
field. While the nonlinear EM-field was shown to incorporate the back reaction into the metric, the multiple 
gauge fields there was shown to shrink the event horizon radius. In the regime, the reduced horizon radius was shown
to be due to a large number of charges arising out of all the gauge fields in the frame-work. The emerging
notion of a $2D$ extremal monopole black hole governing a dS geometry was shown in the regime. On the other hand,
the event horizon of the extremal AdS there transformed to a curvature singularity in dS space. The effective
gravitational potential associated with the reduced mass of extremal dS black hole was analyzed to confirm the presence of three extra large dimensions in the regime. The Hagedron transitions in the near horizon geometry
of $dS_5$ monopole black hole was analyzed to decouple the $\Theta$-terms. The new $dS_5$ vacuum with a nontrivial
cosmological potential possibly describes our brane-world, $i.e.$ a $D_3$-brane, at its boundary. The potential
was argued to be at its local minima on the brane-world  and describes a small positive constant $\Lambda$.

\par
\sp

Finally, a careful analysis may reveal that two different topologies representing $dS_2$ and $AdS_2$ geometries 
are interchanged, $i.e.\ \Re \times S^1 \leftrightarrow S^1\times \Re^1$ in the gravity decoupling regime.
Intuitively, it incorporates an interchange of $2D$ geometries, such as {\it hyperbolic} $\leftrightarrow$
{\it cylinrdrical}. The change in topology is significant  to an emerging two dimensional aspect of space-time 
within a $4D$ effective string theory. It provides an evidence to the notion of signature change discussed
in a collaboration with Majumdar \cite{verlinde2}. Nevertheless, the topology of the $4D$ effective spacetime 
remains unchanged. The tunneling between $dS_2$ and $AdS_2$ vacua is a potential candidate and 
may possess deeper implications in quantum gravity. Though, it may be illuminating to view 
the tunneling analogous to the established {\it closed} $\leftrightarrow$ {\it open} 
string duality, it remains to explore a concrete geometric relation in higher dimensions.

\section*{Acknowledgments}

I would like to thank the Abdus Salam I.C.T.P, Trieste for hospitality, where a part of this work
was completed. The research was supported, in part, by the D.S.T, Govt.of India, under SERC fast track 
young scientist project PSA-09/2002.

\def\anp{Ann. of Phys.}
\def\cmp{Commun. Math. Phys.}
\def\atmp{Adv.Theor.Math.Phys.}
\def\prl{Phys. Rev. Lett.}
\def\prd#1{{Phys. Rev.} {\bf D#1}}
\def\jhep{J.High Energy Phys.}
\def\cqg{{Class. \& Quantum Grav.}}
\def\plb#1{{Phys. Lett.} {\bf B#1}}
\def\npb#1{{Nucl. Phys.} {\bf B#1}}
\def\mpl#1{{Mod. Phys. Lett} {\bf A#1}}
\def\ijmpa#1{{Int. J. Mod. Phys.} {\bf A#1}}
\def\ijmpd#1{{Int. J. Mod. Phys.} {\bf D#1}}
\def\rmp#1{{Rev. Mod. Phys.} {\bf 68#1}}


\end{document}